\documentclass{PoS}
\usepackage{amssymb,amsmath,wrapfig}
\usepackage{graphicx} 
\usepackage{dsfont}
\usepackage{bbold}
\usepackage{cancel}
\usepackage{url}

\DeclareMathOperator{\Tr}{Tr}

\graphicspath{{Figures/}}

\title{Canonical simulations of supersymmetric SU$(N)$ Yang-Mills quantum mechanics}

\ShortTitle{Canonical simulations of SYM quantum mechanics}

\author{Georg Bergner, Hang Liu and \speaker{Urs Wenger} \\
        Albert Einstein Center for Fundamental Physics\\
         Institute for Theoretical Physics\\
        University of Bern\\
        Sidlerstrasse 5\\
        CH--3012 Bern\\
        Switzerland\\
        E-mail:  \email{bergner@itp.unibe.ch}, \email{liu@itp.unibe.ch}, \email{wenger@itp.unibe.ch}}

\abstract{
The fermion loop formulation naturally separates partition functions into their canonical sectors. Here we discuss various strategies to make use of this for supersymmetric SU($N$) Yang-Mills quantum mechanics obtained from dimensional reduction in various dimensions and present numerical results for the separate canonical sectors with fixed fermion numbers. We comment on potential problems due to the sign of the contributions from the fermions and due to flat directions. 
}

\FullConference{The 33rd International Symposium on Lattice Field
  Theory\\
 14 -18 July  2015\\
 Kobe International Conference Center, Kobe, Japan}

\begin{document}

\section{Introduction}

Since long it is suspected that SU($N$) gauge theories can be regarded
as the low energy effective theory of $N$ D-branes in specific
parameter regimes. In this way, dimensionally reduced large-$N$
supersymmetric Yang-Mills (SYM) gauge theories might provide a
nonperturbative formulation of the string/M-theory which describes the
dynamics of the D-branes. In particular, the connection between black
$p$-branes and SYM gauge theories in $d=(p+1)$ dimensions allows to
study black hole thermodynamics through the corresponding strongly
coupled gauge theory.
Apart from providing tests of the gauge/gravity duality, SYM quantum
mechanics is interesting {\it per se}. There are various intriguing
expectations concerning the behaviour of the theory in specific
sectors of fixed fermion number
\cite{Wosiek:2002nm,Campostrini:2004bs}. In particular, while certain
canonical sectors only have a discrete energy spectrum, some sectors
allow for the discrete spectrum to be immersed in a continuous one
reaching down to zero energy. These continuous spectra can presumably
be associated with the so-called flat directions of the potential
which are present in the classical theory, but may or may not survive
the quantisation of the theory.

In these proceedings we report on our ongoing effort to perform
nonperturbative calculations in SYM quantum mechanics with gauge group
SU($N$). Here we concentrate on ${\cal N}=4$ SYM quantum mechanics
which is obtained from dimensional reduction of the $\mathcal{N}=1$
SYM gauge theory in $d=4$ dimensions by compactifying the three
spatial dimensions. To define the theory we employ the lattice
regularisation in Euclidean time proposed in
\cite{Catterall:2007fp,Catterall:2008yz}.  The bosonisation of the
theory on the lattice by means of the fermion loop formulation
\cite{Baumgartner:2014nka,Baumgartner:2015qba,Baumgartner:2015zna}
decomposes the fermion contributions into fermion sectors with fixed
fermion number. Our recent progress in understanding the algebraic
structure of the fermion loop formulation allows the explicit
construction of transfer matrices \cite{Steinhauer:2014oda}. The
transfer matrices in turn provide the starting point for the
construction of local fermion update algorithms which allow to
directly simulate fixed canonical sectors of the theory.

In the following we summarise the derivation of the transfer matrices
for generic fermion number sectors, recapitulating our results from
\cite{Steinhauer:2014oda}, and show first results for some simple
observables such as the Polyakov loop, the moduli of the bosonic
fields and the fermion action from simulations in fixed canonical
sectors of the theory.

\section{Lattice regularisation and canonical sectors}
We directly start with $\mathcal{N}=4$ SYM quantum mechanics obtained
from dimensional reduction of $\mathcal{N}=1$ SYM in $d=4$ dimensions
down to $d=1$ dimension. The action in Euclidean time can be written
as
\begin{equation}
S=\frac{1}{g^2}\int_{0}^{\beta} dt \, \text{Tr} \left\lbrace  \left(  D_t X_i   \right)^2 -\frac{1}{2} \left[  X_i , X_j  \right]^2 
+\overline{\psi} D_t \psi -\overline{\psi} \sigma_i \left[ X_i , \psi \right]
\right\rbrace 
\end{equation}
where $D_t=\partial_t-i[A(t),\cdot]$ denotes the covariant derivative
using the time component of the SU($N$) gauge field $A(t)$, while the
spatial components become the bosonic fields $X_i(t)$ with $i=1, 2,
3$. The anticommuting 2-component complex fermion fields
$\overline{\psi}(t),\, \psi(t)$ interact with the bosonic fields $X_i$
through a Yukawa-type interaction involving the three Pauli-matrices
$\sigma_i$. We note that all the fields are in the adjoint
representation of SU($N$). The discretisation of the Lagrangian on a
time lattice with $L_t$ points is straightforward and yields for the
bosonic part
\begin{equation}
 {S}_B = \frac{1}{g^2}\sum_{t=0}^{L_t-1} \, \Tr \left\lbrace  {D}_t X_i(t) {D}_t X_i(t)   -\frac{1}{2} \left[  X_i(t) , X_j(t)  \right]^2 \right\rbrace
\label{eq:discretised boson action}
\end{equation}
where ${D}_t X_i(t)= U(t) X_i(t+1) U^\dagger(t)-X_i(t)$ is the
covariant forward derivative and $U(t)$ is an element of the gauge
group SU($N$). For the fermionic part of the action we introduce a
Wilson term with Wilson parameter $r=1$ in order to avoid fermion
doubling. With this choice of $r$ the massless Wilson Dirac operator
in one dimension involves just the forward derivative and one obtains
\begin{equation}
{S}_F=\frac{1}{g^2}\sum_{t=0}^{L_t-1} \, \Tr \left\lbrace \overline{\psi}(t) D_t \psi(t) -\overline{\psi}(t) \sigma_i \left[ X_i(t) , \psi(t) \right] \right\rbrace 
\end{equation}
for the fermion action. More specifically, we have
\begin{equation}
{S}_F=\frac{1}{2g^2} \sum_{t=0}^{L_t-1} \, \left[
  -\overline{\psi}_\alpha^a(t) W^{ab}_{\alpha\beta}(t) e^{+\mu} \psi_\beta^b(t+1) 
							       +\overline{\psi}_\alpha^a(t)\Phi_{\alpha \beta}^{ac}(t)\psi_\beta^c(t)  \right] \equiv \overline{\psi} {\cal D}_{p,a}[U,X_i;\mu] \psi
\label{eq:discretised fermion action}
\end{equation}
where $W(t)$ denote the real adjoint gauge link matrices
\begin{equation}
W^{ab}_{\alpha\beta}(t) = 2 (\sigma_0)_{\alpha\beta} \otimes \Tr\{T^a U(t) T^b U(t)^\dagger\} \, .
\end{equation}
Note that we have introduced a chemical potential $\mu$ in the
standard way \cite{Hasenfratz:1983ba}.  The subscripts $p, a$ for the
Dirac matrix ${\cal D}$ denote periodic or antiperiodic boundary
conditions, respectively, for the fermionic fields.
The Yukawa interaction matrices $\Phi(t)$ are $n_f^\text{max} \times
n_f^\text{max}$ with $n_f^\text{max}=2 (N^2-1)$ and read
\begin{equation}
 \Phi_{\alpha \beta}^{ac}(t) = (\sigma_0)_{\alpha\beta} \otimes \delta^{ac} -
 2 \, (\sigma_i)_{\alpha\beta} \otimes \Tr\{T^a [X_i(t), T^c]\} \, .
\end{equation}

Two remarks are in order. Firstly, all supersymmetry breaking terms
apart from the ones introduced by the discretisation in
eq.~(\ref{eq:discretised boson action}) and (\ref{eq:discretised
  fermion action}) are forbidden by the gauge symmetry. Hence,
supersymmetry is expected to be automatically restored in the
continuum limit without any fine tuning
\cite{Catterall:2007fp,Catterall:2008yz}. Secondly, the Wilson term
breaks the time reversal symmetry, or equivalently the particle-hole
exchange symmetry. This reflects itself in the fact that the action in
eq.~(\ref{eq:discretised fermion action}) only allows forward
propagating fermions. As a consequence, the exchange symmetry between
the related fermion sectors with $n_f$ and $n_f^\text{max} - n_f$
becomes exact only in the continuum limit.

Let us now derive exact expressions for the fermionic contributions to
the partition function of the theory for a given fixed gauge and boson
field background -- the canonical determinants. In quantum mechanics
the lattice regulated determinant of the Dirac matrix can readily be
calculated, and one obtains
\begin{equation}
\det {\cal D}_{p,a}[U,X_i;\mu] = \det \left[{\cal T} \mp e^{+\mu L_t} \right] \quad \text{with} \quad {\cal T} \equiv \prod_{t=0}^{L_t-1} \Phi(t) W(t) \, .
\end{equation}
This essentially corresponds to the dimensionally reduced determinant
for Wilson fermions derived in \cite{Alexandru:2010yb,Nagata:2010xi}
except that here the dimensional reduction is from the full matrix to
a $n_f^\text{max}\times n_f^\text{max}$ 'flavour' matrix. It is now
easy to get the canonical determinants from the fugacity expansion
\begin{equation}
\det {\cal D}_{p,a}[U,X_i;\mu] =  \sum_{n_f=0}^{2(N^2-1)} (\mp e^{\mu L_t})^{n_f}
\det{\cal D}_{n_f}[U,X_i] 
\end{equation}
which identifies the canonical determinants as the coefficients of the
characteristic polynomial. These coefficients can be expressed in
terms of the elementary symmetric functions $S_k$ of order $k$ of the
eigenvalues $\{\tau_i, i=1,\ldots,n_f^\text{max}\}$ of ${\cal T}$,
\begin{equation}
S_k({\cal T}) \equiv S_k(\{\tau_i\}) = \sum_{1\leq
  i_1 < \cdots < i_k \leq n_f^\text{max}} \prod_{j=1}^k
\tau_{i_j} \, ,
\end{equation}
and one eventually obtains
\begin{equation}
\det {\cal D}_{n_f}[U,X_i] = S_{n_f^\text{max} - n_f}({\cal T}) \, .
\label{eq:detQ symmetric polynomial}
\end{equation}
So the crucial object for the calculation of the canonical
determinants is the product ${\cal T}$ of the matrices $\Phi(t)$ and
$W(t)$ which in fact is a product of transfer matrices
\cite{Steinhauer:2014oda}, as was suspected already in
\cite{Alexandru:2010yb}.

\section{Transfer matrices for the canonical sectors}
The explicit construction of the fermion transfer matrices for each
fermion sector is most easily done via the fermion loop formulation
\cite{Steinhauer:2014oda} which in essence is an exact (hopping)
expansion of the fermionic Boltzmann factor to all orders. In this
formulation, the contributions of the fermions to the partition
function are obtained by summing over all possible closed oriented
fermion loops which are forward propagating in time for any given
gauge and boson field background. The loop configuration space
naturally separates into subspaces characterised by the number of
forward propagating fermions $n_f$. The transfer matrix elements are
explicitly given in terms of the cofactors $C(\Phi)$ and the
complementary minors $M(W)$,
\begin{eqnarray}
  \left(T^\Phi_{n_f}\right)_{AB} &=& C_{\bcancel{B}\bcancel{A}}(\Phi)  \,\,\,\, =  (-1)^{p(A,B)} \det  \Phi^{\bcancel{B}\bcancel{A}} \, , \\
\left(T^W_{n_f}\right)_{AB} &=&  M_{AB}(W)  
= \det W^{AB} \, ,
\end{eqnarray}
where $A, B$ are sets of indices $A, B \subseteq
\{1,\ldots,n_f^\text{max}\}$ of order $n_f$ and $p(A,B)=\sum_{i\in A}
i + \sum_{j\in B}j$. $\Phi^{\bcancel{B}\bcancel{A}}$ denotes the
matrix obtained from $\Phi$ by deleting the rows with indices from $B$
and the columns with indices from $A$, while $W^{AB}$ denotes the
matrix obtained from $W$ by picking only the rows with indices from
$A$ and columns with indices from $B$. The size of the transfer
matrices is given by the number of such sets for a given $n_f$ and
corresponds to the number of forward propagating fermion states
$N_\text{states} = n_f^\text{max}!/(n_f^\text{max}-n_f)! \cdot
n_f!$. The fermion contribution to the partition function in sector
$n_f$ is then simply given by
\begin{equation}
\det {\cal D}_{n_f}[U,X_i] = \Tr\left[ \prod_{t=0}^{L_t-1}T^{\Phi}_{n_f}(t)\cdot T^W_{n_f}(t) \right]
\end{equation}
and one can use the Cauchy-Binet formula and some further algebra
\cite{Steinhauer:2014oda} to show that
\begin{equation}
\left[\prod_{t=0}^{L_t-1} T_{n_f}^\Phi(t) \cdot T_{n_f}^W(t)
\right]_{AB} = (-1)^{p(A,B)} \det {\cal T}^{\bcancel{A}\bcancel{B}} = C_{\bcancel{A}\bcancel{B}}({\cal T}) \, ,
\end{equation}
hence the canonical determinant is simply given by the sum over the
principal minors of order $n_f$ of ${\cal T}$ denoted by $
E_{n_f}({\cal T})$,
\begin{equation}
\det {\cal D}_{n_f}[U,X_i] = \sum_{B} \det {\cal T}^{\bcancel{B}\bcancel{B}}  \equiv E_{n_f}({\cal T}) \, .
\label{eq:detQ principal minors}
\end{equation}
Finally, it is easy to show that $E_{n_f}({\cal T}) =
S_{n_f^\text{max}-n_f}({\cal T})$ which proves the equivalence between
the representation using the transfer matrices and the one in
eq.~(\ref{eq:detQ symmetric polynomial}).

Some remarks are in order. Firstly, we note that the matrix ${\cal T}$
describes the dimensionally reduced effective action for the Polyakov
loop coupled to the bosonic fields $X_i$. Secondly, our result for the
canonical determinants in principle allows for local fermion update
algorithms, but in practice only the sectors with $n_f=0$ and
$n_f=n_f^\text{max}$ can be implemented straightforwardly, while in
other sectors algorithms along the lines in \cite{Wenger:2008tq} can
be employed. Thirdly, the construction of the transfer matrices and
the calculation of the canonical determinants in terms of those is
applicable to QCD, since the algebraic structures of the theories are
the same.

\section{Canonical simulations}
Here we present our first results from simulations of the system
employing the gauge group SU($N$) with $N=3$ and $n_f^\text{max} = 16$
directly in the various canonical sectors\footnote{Note also the
  recent effort using the RHMC algorithm for SU(2) in
  \cite{Ambrozinski:2014oka}.}. First we note that the canonical
determinants are real because the eigenvalues $\tau_i$ of ${\cal T}$
are real or come in complex conjugate pairs.
\begin{figure}[t]
\includegraphics[width=0.49\textwidth]{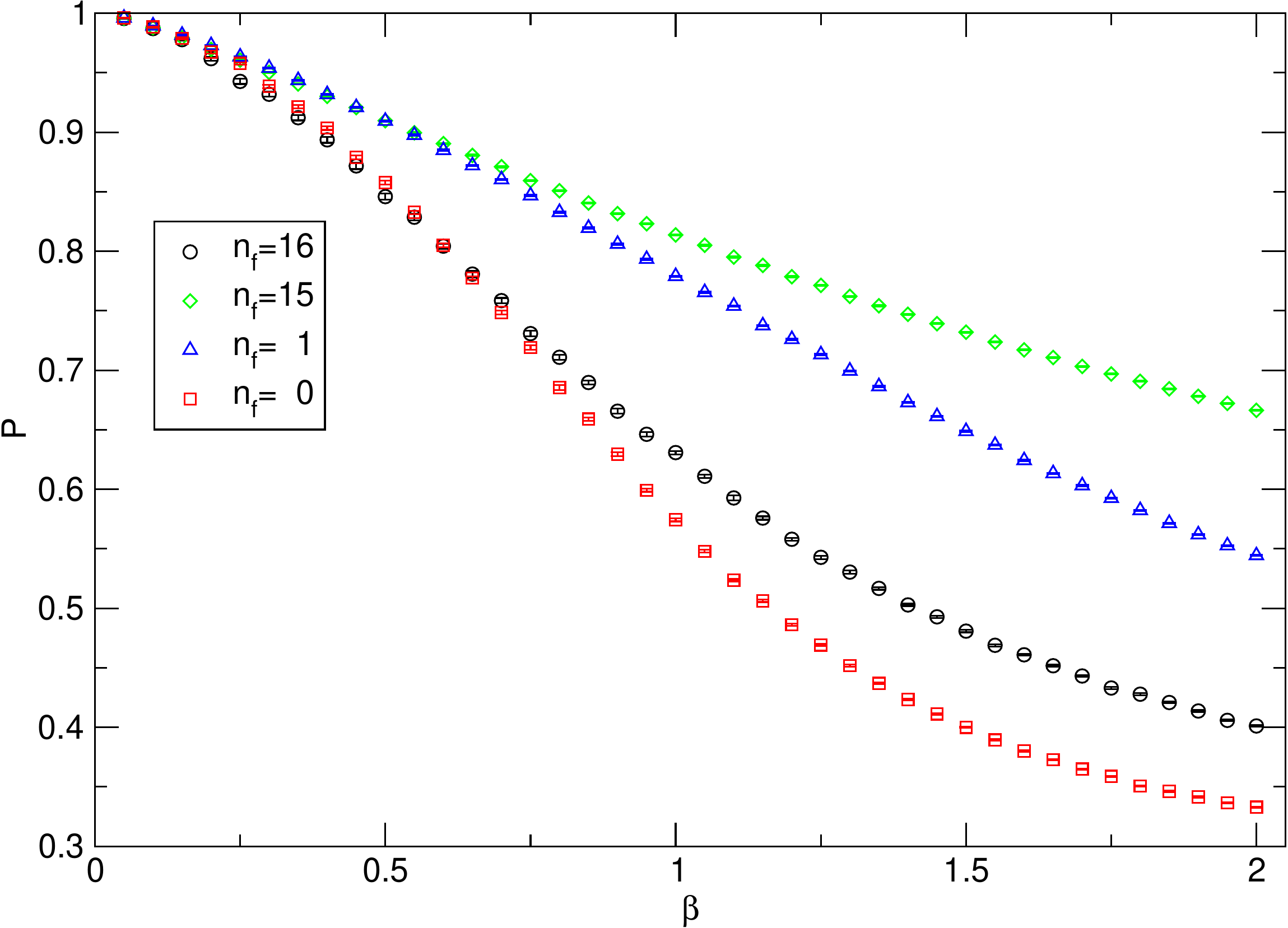}\,\,\,
\includegraphics[width=0.49\textwidth]{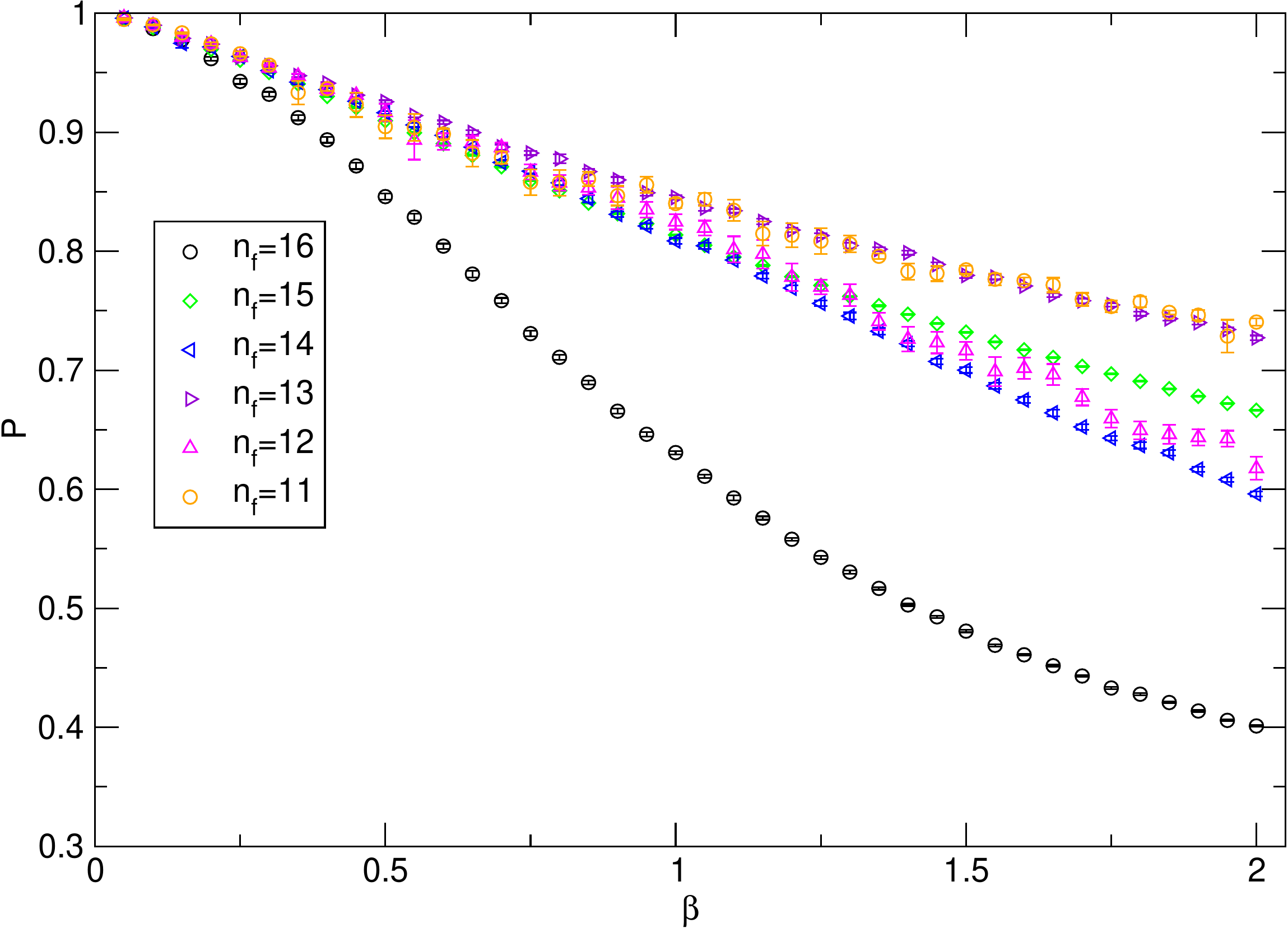}
\caption{The modulus $P$ of the Polyakov loop as a function of $\beta$
  for SU$(3)$ on a $L_t=5$ lattice in various canonical sectors.}
\label{fig:polyakovLoop}
\end{figure}
Furthermore, for the sectors $n_f = 0$ and $n_f=n_f^\text{max}$
(quenched) one can prove that the canonical determinants are
positive. In these two sectors we update the bosonic degrees of
freedom using a local Metropolis algorithm. In the other sectors we
use Metropolis updates based on eq.~(\ref{eq:detQ principal minors})
and currently simulate only in the configuration space with positive
determinants.

In the following we show results on a lattice with temporal extent
$L_t=5$. We measure the moduli of the Polyakov loop and the scalar
field defined by
\begin{equation}
P = \left|\text{Tr} \, \prod_t U(t)\right|, \quad\quad\quad R^2
\equiv |X|^2 = X_i^a X_i^a \, .
\end{equation}
We note that in some sectors the simulations become unstable and $R^2$
grows without bound. We believe that this is because the flat
directions become unstable due to lattice artefacts, and we expect the
behaviour to disappear towards the continuum limit.  In the left panel
of Figure \ref{fig:polyakovLoop} we show $P$ as a function of the
temporal extent of the system parametrised by $\beta$ for the sectors
$n_f=0$ and 16, and for $n_f=1$ and 15. Each pair should be degenerate
in the continuum and we see that this is indeed the case for $\beta
\lesssim 0.6$, while for larger values of $\beta$ lattice artefacts
lift the degeneracy. The (physical) differences between the sectors is
illustrated in the right panel of Figure \ref{fig:polyakovLoop} where
we show $P$ in the sectors $n_f=16$ down to $11$.

\begin{figure}[t]
\includegraphics[width=0.49\textwidth]{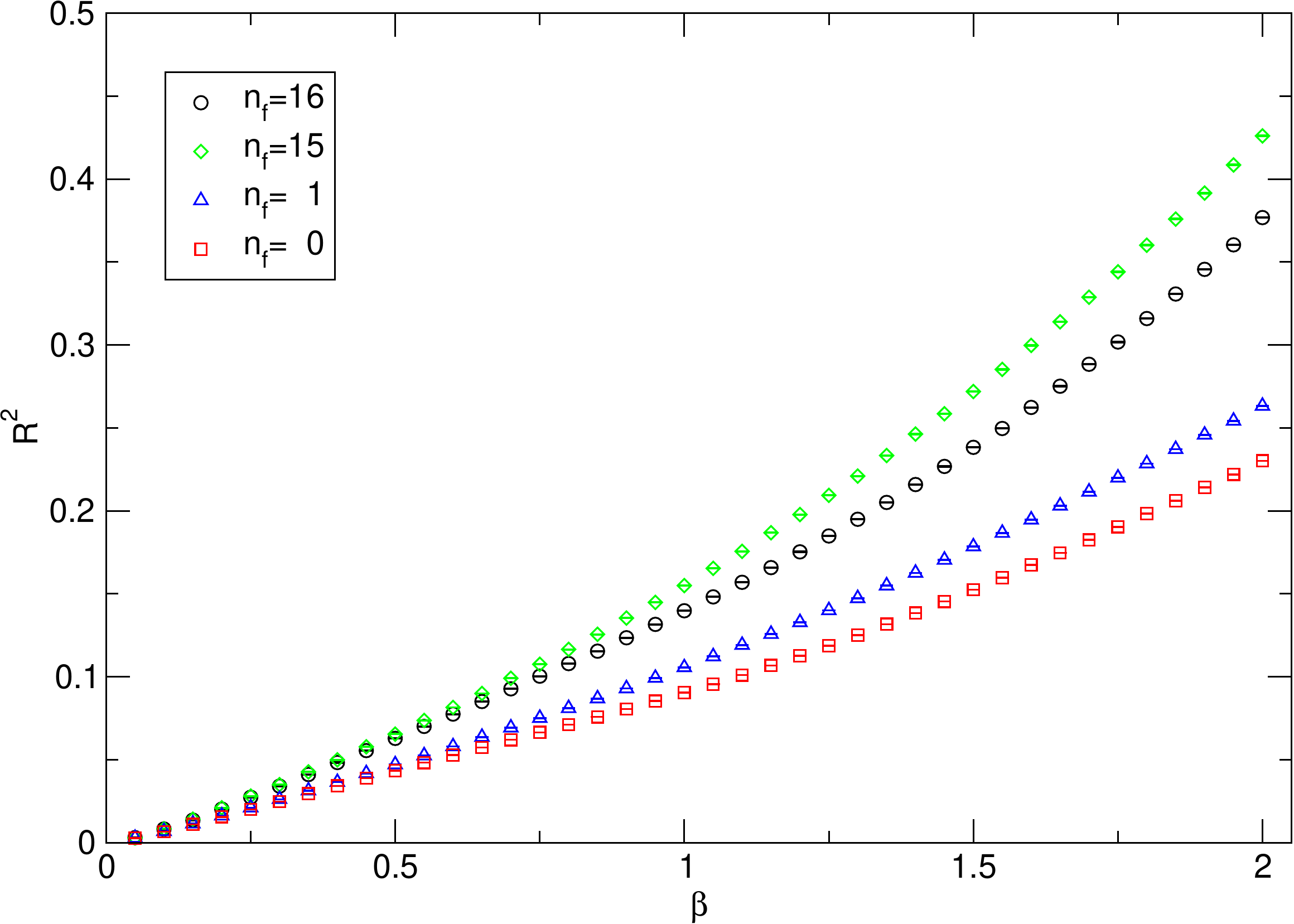}\,\,\,
\includegraphics[width=0.49\textwidth]{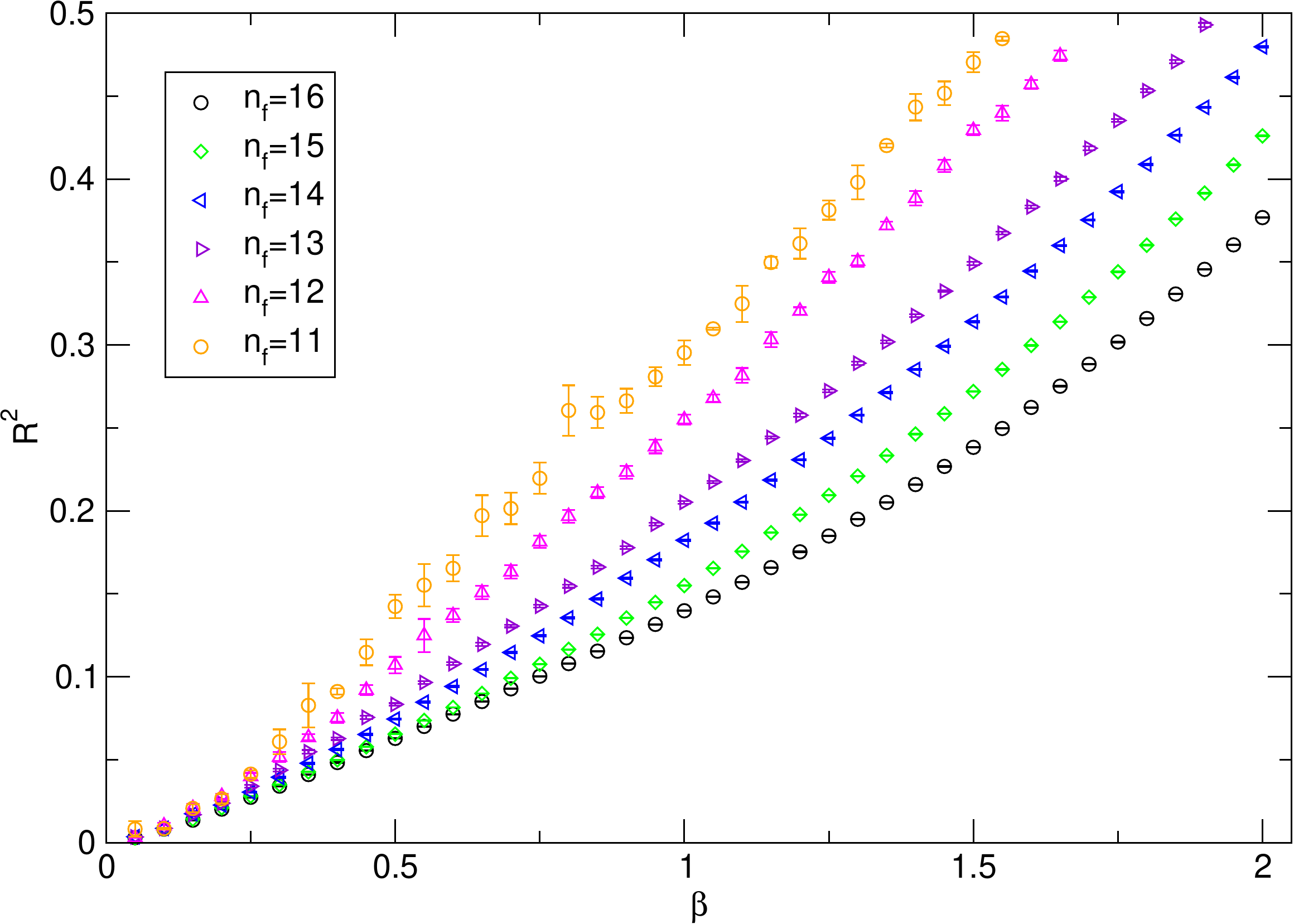}
\caption{The square of the modulus of the scalar field $R^2$ as a
  function of $\beta$ for SU$(3)$ on a $L_t=5$ lattice in various
  canonical sectors.}
\label{fig:moduliX2}
\end{figure}

Next, in Figure \ref{fig:moduliX2} we show the square of the modulus
of the scalar field $R^2$ as a function of $\beta$ for the same
combinations of sectors as in Figure \ref{fig:polyakovLoop}. Again we
find that the degeneracy between the $n_f=0$ and 16, $n_f=1$ and 15,
and so on, is lifted by lattice artefacts towards large values of
$\beta$.

\begin{figure}[b]
\centering
\includegraphics[width=0.575\textwidth]{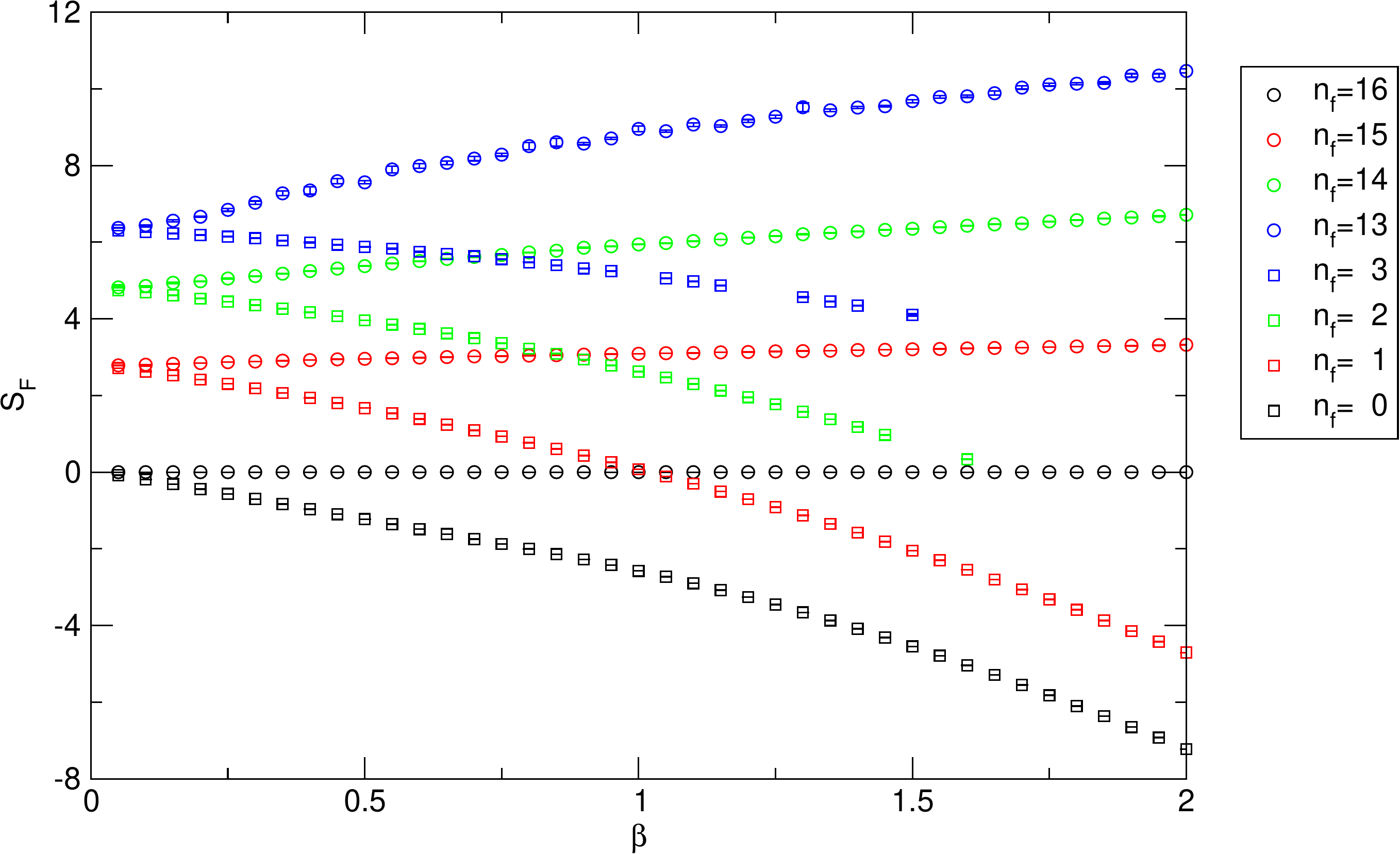}
\caption{The fermionic action as a function of $\beta$ for SU$(3)$ on
  a $L_t=5$ lattice in various canonical sectors.}
\label{fig:lnDet}
\end{figure}

Finally, in Figure \ref{fig:lnDet} we show the fermionic action
$S_F=\langle \text{ln det }{\cal D}_{n_f} \rangle_{n_f}$ as a function
of $\beta$ in various canonical sectors. We find that the degeneracy
of this observable between the mirror sectors becomes better and
better towards $\beta \rightarrow 0$, suggesting that reweighting
between the mirror sectors could become feasible in that limit, or
more generally towards the continuum limit.

\section{Summary and outlook}
In this contribution we summarise the derivation of explicit transfer
matrices for ${\cal N}=4$ SYM quantum mechanics with generic gauge
group SU($N$) discretised on a time lattice. The transfer matrices are
defined separately in each canonical sector with fixed fermion number
$n_f$ and form the basis for canonical simulations of the theory. One
caveat is that in those sectors where the canonical determinants are
not positive definite, the local Metropolis algorithm is currently not
very efficient and only samples configurations with positive
determinants.

Several paths are now open for further investigation. From an
algorithmic viewpoint, it is interesting to examine the systematics of
reweighting ensembles of configurations from one fermion sector to
another, or from simulations at finite (imaginary) chemical
potential. Concerning the physics of the model, it is interesting to
calculate correlation functions and energy spectra in the various
canonical sectors. The investigation of the phase transition in the
large-$N$ limit of the ${\cal N}=16$ SYM quantum mechanics is most
useful for a further understanding of the thermodynamics of certain
black holes. The results in these proceedings are a first step towards
these goals.

\bibliography{csoSYMqm_proceedings}

\end{document}